\documentclass[aps,prl,twocolumn,superscriptaddress,showpacs,preprintnumbers,amsmath,amssymb]{revtex4}
\usepackage{graphicx} 
\usepackage{dcolumn}  
\usepackage{amsmath}
\usepackage{epsfig}
\usepackage{xspace}
\usepackage{multirow}
\usepackage{color}
\usepackage{ulem}     

\definecolor{Bisque}{rgb}{.996,.891,.755}
\definecolor{Red}{rgb}{.996,.0,.0}
\definecolor{Blue}{rgb}{.0,.746,.996}
\definecolor{Green}{rgb}{.0,.996,.496}
\definecolor{Magenta}{rgb}{.996,.0,.996}
\def\kstarbar{\overline{K}{}^{*0}}
\def\dbar{\overline{D}{}^{0}}
\def\bbar{\overline{B}{}^{0}}

\renewcommand{\arraystretch}{1.1}
\def\myspecial#1{}                   

\begin{document}

  
\title{ \quad\\[1.0cm] \Large 
Search for
$B^0 \to K^{*0}\kstarbar$, $B^0 \to K^{*0} K^{*0}$ and 
$B^0 \to K^+\pi^- K^{\mp}\pi^{\pm}$ Decays
}

\tighten

\affiliation{Budker Institute of Nuclear Physics, Novosibirsk}
\affiliation{Faculty of Mathematics and Physics, Charles University, Prague}
\affiliation{University of Cincinnati, Cincinnati, Ohio 45221}
\affiliation{Justus-Liebig-Universit\"at Gie\ss{}en, Gie\ss{}en}
\affiliation{Hanyang University, Seoul}
\affiliation{University of Hawaii, Honolulu, Hawaii 96822}
\affiliation{High Energy Accelerator Research Organization (KEK), Tsukuba}
\affiliation{Institute of High Energy Physics, Vienna}
\affiliation{Institute of High Energy Physics, Protvino}
\affiliation{Institute of Mathematical Sciences, Chennai}
\affiliation{Institute for Theoretical and Experimental Physics, Moscow}
\affiliation{J. Stefan Institute, Ljubljana}
\affiliation{Kanagawa University, Yokohama}
\affiliation{Institut f\"ur Experimentelle Kernphysik, Karlsruhe Institut f\"ur Technologie, Karlsruhe}
\affiliation{Korea University, Seoul}
\affiliation{Kyungpook National University, Taegu}
\affiliation{\'Ecole Polytechnique F\'ed\'erale de Lausanne (EPFL), Lausanne}
\affiliation{University of Maribor, Maribor}
\affiliation{Max-Planck-Institut f\"ur Physik, M\"unchen}
\affiliation{University of Melbourne, School of Physics, Victoria 3010}
\affiliation{Nagoya University, Nagoya}
\affiliation{Nara Women's University, Nara}
\affiliation{National Central University, Chung-li}
\affiliation{National United University, Miao Li}
\affiliation{Department of Physics, National Taiwan University, Taipei}
\affiliation{H. Niewodniczanski Institute of Nuclear Physics, Krakow}
\affiliation{Nippon Dental University, Niigata}
\affiliation{Niigata University, Niigata}
\affiliation{University of Nova Gorica, Nova Gorica}
\affiliation{Novosibirsk State University, Novosibirsk}
\affiliation{Osaka City University, Osaka}
\affiliation{Panjab University, Chandigarh}
\affiliation{University of Science and Technology of China, Hefei}
\affiliation{Seoul National University, Seoul}
\affiliation{Sungkyunkwan University, Suwon}
\affiliation{School of Physics, University of Sydney, NSW 2006}
\affiliation{Tata Institute of Fundamental Research, Mumbai}
\affiliation{Excellence Cluster Universe, Technische Universit\"at M\"unchen, Garching}
\affiliation{Toho University, Funabashi}
\affiliation{Tohoku Gakuin University, Tagajo}
\affiliation{Tohoku University, Sendai}
\affiliation{Department of Physics, University of Tokyo, Tokyo}
\affiliation{Tokyo University of Agriculture and Technology, Tokyo}
\affiliation{IPNAS, Virginia Polytechnic Institute and State University, Blacksburg, Virginia 24061}
\affiliation{Yonsei University, Seoul}
  \author{C.-C.~Chiang}\affiliation{Department of Physics, National Taiwan University, Taipei} 
  \author{H.~Aihara}\affiliation{Department of Physics, University of Tokyo, Tokyo} 
  \author{K.~Arinstein}\affiliation{Budker Institute of Nuclear Physics, Novosibirsk}\affiliation{Novosibirsk State University, Novosibirsk} 
  \author{V.~Aulchenko}\affiliation{Budker Institute of Nuclear Physics, Novosibirsk}\affiliation{Novosibirsk State University, Novosibirsk} 
  \author{T.~Aushev}\affiliation{\'Ecole Polytechnique F\'ed\'erale de Lausanne (EPFL), Lausanne}\affiliation{Institute for Theoretical and Experimental Physics, Moscow} 
  \author{A.~M.~Bakich}\affiliation{School of Physics, University of Sydney, NSW 2006} 
  \author{V.~Balagura}\affiliation{Institute for Theoretical and Experimental Physics, Moscow} 
  \author{E.~Barberio}\affiliation{University of Melbourne, School of Physics, Victoria 3010} 
  \author{K.~Belous}\affiliation{Institute of High Energy Physics, Protvino} 
  \author{V.~Bhardwaj}\affiliation{Panjab University, Chandigarh} 
% ...
\collaboration{The Belle Collaboration}


\begin{abstract}
We report a search for the decays 
$B^0\to K^{*0} \kstarbar$ and $B^0\to K^{*0} K^{*0}$. 
We also measure other charmless decay modes with 
$K^+\pi^-K^-\pi^+$ and $K^{+}\pi^{-}K^{+}\pi^{-}$ final states. 
The results are obtained from a data sample containing 
$657 \times 10^6$ $B \overline B$ pairs collected with the Belle 
detector at the KEKB asymmetric-energy $e^+e^-$ collider. 
We set upper limits on the branching fractions for
$B^0\to K^{*0} \kstarbar$ and 
$B^0\to K^{*0} K^{*0}$ of $0.81 \times 10^{-6}$ and 
$0.20\times 10^{-6}$, respectively, at the 90\% confidence level. 
\end{abstract}

\pacs{11.30.Er, 12.15.Hh, 13.25.Hw, 14.40.Nd}

\maketitle
\tighten
\normalsize

The charmless decay $B^0\to K^{*0}\kstarbar$~\cite{chargeconjugates}
proceeds through electroweak and gluonic $b \to d$ 
``penguin'' loop diagrams. It provides an opportunity to 
probe the dynamics of both weak and strong interactions, 
which play an important role in 
$CP$ violation phenomena.
For a $B$ meson decaying to two vector particles, $B\to VV$, 
theoretical models based on the frameworks of either 
QCD factorization or perturbative QCD predict 
the fraction of longitudinal 
polarization ($f_{\mathrm{L}}$) to be $\sim 0.9$ for 
tree-dominated decays and $\sim 0.75$ for 
penguin-dominated decays~\cite{ali, chchen}. 
However, the measured polarization fraction in the 
pure penguin decay $B \to \phi K^*$ has a somewhat 
lower value of $f_{\mathrm{L}} \sim 0.5$~\cite{kfchen}. 
This unexpected result has motivated further studies~\cite{kagan}. 

One resolution to this puzzle is a smaller $B\to K^*$ 
form factor that could reduce $f_{\mathrm{L}}$ significantly \cite{hnLi}. 
If this explanation is correct, the penguin-dominated decay 
$B^0\to K^{*0} \kstarbar$ should exhibit a similar 
polarization fraction. A time-dependent angular analysis of 
$B^0\to K^{*0} \kstarbar$ could distinguish between 
penguin annihilation and rescattering as mechanisms for the 
$f_{\mathrm{L}}$ observed in $B \to \phi K^*$ decays \cite{datta}. 
The $B^0\to K^{*0} \kstarbar$ mode can also be used to 
extract the 
branching fraction corresponding to 
the longitudinal helicity final state, 
determine hadronic parameters for the $b\to s$ decay 
$B_s \to K^{*0} \kstarbar$, and help constrain 
the angles $\phi_2 \ (\alpha)$ and $\phi_3 \ (\gamma)$ 
of the Cabibbo-Kobayashi-Maskawa unitarity triangle 
\cite{atwood}.
The topologically similar decay $B^0\to K^{*0} K^{*0}$ 
is strongly suppressed in the Standard Model (SM); 
its observation would indicate new physics.

Theoretical calculations predict the branching fractions 
for $B^0\to K^{*0} \kstarbar$ and $B^0\to K^{*0} K^{*0}$ 
to be $(0.17-0.92) \times 10^{-6}$~\cite{aali} 
and $(2.9\pm 0.2)\times 10^{-15}$~\cite{DanP}, respectively.
The BaBar collaboration \cite{aaubert} has reported
a branching fraction 
$\mathcal{B}(B^0\to K^{*0} \kstarbar) = 
(1.28^{+0.35}_{-0.30} \pm 0.11)\times 10^{-6}$ 
and has set an upper limit 
$\mathcal{B}(B^0\to K^{*0} K^{*0}) < 0.41 \times 10^{-6}$
at the 90\% 
confidence level (C.L.).
In this paper, we report a search for the decays 
$B^0\to K^{*0} \kstarbar$, $B^0\to K^{*0} K^{*0}$, and 
other charmless decay modes with a $K^+\pi^-K^-\pi^+$ or 
$K^{+}\pi^{-}K^{+}\pi^{-}$ final state using a data sample 
1.7 times larger than that of BaBar.
The data sample used in the analysis 
contains $657 \times 10^6$ $B \overline B$ pairs 
collected with the 
Belle detector~\cite{102, 103} 
at the KEKB asymmetric-energy $e^+e^-$ 
(3.5~GeV and 8~GeV) collider~\cite{101} 
operating at the $\Upsilon(4S)$ resonance. 
The Belle detector includes a silicon vertex detector, 
a 50-layer central drift chamber (CDC), an array of 
aerogel threshold Cherenkov counters (ACC), and a 
barrel-like arrangement of time-of-flight scintillation 
counters (TOF). 
Signal Monte Carlo (MC) events are generated with EVTGEN \cite{Lange}, 
and final-state radiation is taken into account with the PHOTOS 
package \cite{105}. Generated events are processed through a 
full detector simulation program based on GEANT3 \cite{Brun}.

We reconstruct signal decays from neutral 
combinations of four charged tracks fitted to a common vertex.
Neutral $K^*$ mesons are reconstructed via $K^{*0} \to K^+\pi^-$ 
and $\kstarbar \to K^-\pi^+$. 
Charged track candidates are required to have a 
distance of closest approach to the interaction 
point of less than 2.0 cm in the direction along 
the positron beam ($z$ axis), and less than 0.1 cm in 
the transverse plane. Tracks are also required to have
a laboratory momentum in the range [0.5, 4.0] GeV/$c$, 
a polar angle in the range $[32.2, 127.2]^{\circ}$, and 
a transverse momentum $p_T>0.1$ GeV/$c$. 
Charged pions are identified using information from the 
CDC ($dE/dx$), ACC, and TOF detectors~\cite{ENakano}.
We distinguish charged kaons from pions using a likelihood ratio 
$\mathcal{R}_{K}=\mathcal{L}_{K}/(\mathcal{L}_{K}+
\mathcal{L}_{\pi})$, where $\mathcal{L}_{\pi}(\mathcal{L}_{K})$ 
is the likelihood value for the pion (kaon) hypothesis. 
We require $\mathcal{R}_{K} < 0.4$ ($\mathcal{R}_{K} > 0.6$) 
for the two charged pions (kaons). 
The kaon (pion) identification efficiency is 83\% (90\%), and 
6\% (12\%) of pions (kaons) are misidentified as kaons (pions). 
We also use the lepton identification likelihood $\mathcal{R}_x$ 
($x$ denotes either $\mu$ or $e$) described in Ref.~\cite{ENakano}: 
charged particles identified as electrons ($\mathcal{R}_e > 0.9$) 
or muons ($\mathcal{R}_{\mu} > 0.9$) are removed.

We veto $B\to D^{(*)\pm}X$, $B\to D_s^{\pm} X$, 
$B\to D^0 X$, and $B^0\to \phi X$ decays that result 
in the $K^+\pi^-K^-\pi^+$ final state, 
and we veto $B\to D^{(*)\pm}X$ and 
$B\to D^0 X$ decays that result in the $K^+\pi^-K^+\pi^-$ 
final state. For the $B\to D^{(*)\pm}X$ and $B\to D_s^{\pm} X$ vetos, 
we remove candidates that satisfy either
$|M(K^{\pm}K^{\mp}\pi^{\mp})-m_{D_{(s)}^{\mp}}|<13 \ \mathrm{MeV}/c^2$, 
$|M(K^{\pm}\pi^{\mp}\pi^{\mp})-m_{D_{(s)}^{\mp}}|<13 \ \mathrm{MeV}/c^2$, 
$|M(K^{\pm}h_{K}^{\mp}\pi^{\mp})-m_{D^{\mp}}|<13 \ \mathrm{MeV}/c^2$, 
or $|M(K^{\pm}h_{\pi}^{\mp}\pi^{\mp})-m_{D^{\mp}}|<13 \ \mathrm{MeV}/c^2$,
where $m_{D_{(s)}^{\mp}}$ are the masses of the $D_{(s)}^{\mp}$ mesons, and
$h_{K}^{\mp} \ (h_{\pi}^{\mp})$ is the kaon (pion) mass assigned 
to a pion (kaon) candidate track
[i.e., a kaon (pion) was misidentified as a pion (kaon)].
For the $B\to D^0 X$ veto, we remove candidates satisfying either
$|M(K^{\pm}K^{\mp})-m_{D^0}|<13 \ \mathrm{MeV}/c^2$ or 
$|M(K^{\pm}h_{K}^{\mp})-m_{D^0}|<13 \ \mathrm{MeV}/c^2$,
where $m_{D^0}$ is the mass of the $D^0$ meson. 
For the $B\to \phi X$ veto, we remove candidates satisfying 
$|M(K^{\pm}h_{K}^{\mp})-m_{\phi}|<20 \ \mathrm{MeV}/c^2$, 
where $m_{\phi}$ is the mass of the $\phi$ meson.
These vetos together remove 9.7\% (3.6\%) of 
longitudinally polarized $B^0\to K^{*0} \kstarbar$ 
($B^0\to K^{*0} K^{*0}$) signal, according to MC simulation.

Signal event candidates are characterized by two 
kinematic variables: the beam-energy-constrained mass, 
$M_{\mathrm{bc}}=\sqrt{E^2_{\mathrm{beam}}-P^{*2}_{B}}$, 
and the energy difference, $\Delta E = E^*_{B}-E_{\mathrm{beam}}$, 
where $E_{\mathrm{beam}}$ is the run-dependent beam energy, 
and $P_B^*$ and $E_B^*$ 
are the momentum and energy of the 
$B$ candidate in the $\Upsilon(4S)$ center-of-mass (CM) frame. 
We distinguish nonresonant $B^0\to KK\pi\pi$ decays from our signal
modes by fitting the two-dimensional mass distributions
$M(K^+\pi^-)$ vs.~$M(K^-\pi^+)$ or
$M(K^+\pi^-)$ vs.~$M(K^+\pi^-)$.
There are two possible combinations in 
$B^0 \to K^{*0}K^{*0}$ reconstruction for 
$M(K^+\pi^-)$ vs.\ $M(K^+\pi^-)$: 
$(K^+_1\pi^-_1)(K^+_2\pi^-_2)$ and 
$(K^+_1\pi^-_2)(K^+_2\pi^-_1)$, where the 
subscripts label the momentum ordering, i.e., 
$K^+_1$($\pi^-_1$) has higher momentum than $K^+_2$($\pi^-_2$). 
We consider both $(K^+_1\pi^-_1)(K^+_2\pi^-_2)$ and 
$(K^+_1\pi^-_2)(K^+_2\pi^-_1)$ combinations and select 
candidate events if either one of the combined masses 
lies within the signal window of [0.7, 1.7]~GeV/$c^2$. 
If both combinations fall
within the signal window, we select the
$(K^+_1\pi^-_2)(K^+_2\pi^-_1)$ combination. 
According to MC simulation, this choice selects the 
correct combination for signal decays 99\% of the time. 
For fitting, we symmetrize the $M^2(K^+\pi^-)$ vs.~$M^2(K^+\pi^-)$ 
plot by plotting 
$M^2(K^+_1\pi^-) \ [M^2(K^+_2\pi^-)]$
on the horizontal axis for events with 
an even [odd] event number. 
This number denotes the location of the event in the 
data set (i.e., $n_{\mathrm{event}}=1,2,3...N_{\mathrm{total}}$).

The dominant source of background is continuum 
$e^+e^- \to q\bar{q}$~($q=u,d,s$ and $c$) events. 
To distinguish signal from the jet-like continuum background, 
we use modified Fox-Wolfram moments \cite{104} that are 
combined into a Fisher discriminant. 
This discriminant is subsequently combined with the 
probabilities for the cosine of the $B$ flight direction 
in the CM frame and the distance along the $z$ axis between 
the two $B$ meson decay vertices to form a likelihood ratio 
$\mathcal{R}=\mathcal{L}_{s}/(\mathcal{L}_{s}+\mathcal{L}_{q\overline q})$. 
Here, $\mathcal{L}_{s}$ ($\mathcal{L}_{q\overline q}$) is a likelihood
function for signal (continuum) events that is obtained
from the signal MC simulation (events in the sideband 
region $M_{\rm bc}<5.26$~GeV/$c^2$). 
For additional suppression,
we also use a flavor tagging quality variable $r$ 
provided by the Belle tagging algorithm~\cite{112} that identifies
the flavor of the accompanying $B^0$ meson in the
$\Upsilon(4S)\to B^0\bbar$ decay.
The variable $r$ ranges from $r=0$ for no flavor discrimination 
to $r=1$ for unambiguous flavor assignment, and it is used 
to divide the data sample into six $r$ bins.
As the discrimination between signal and continuum events
depends on the $r$-bin, we impose different requirements
on $\mathcal{R}$ for each bin. The requirements are determined
by maximizing
a figure-of-merit $N_s / \sqrt{N_s + N_{q\overline q}}$,
where $N_s$ $(N_{q\overline q})$ is the expected number of
signal (continuum) events in the signal region
$\Delta E \in [-0.045,\ 0.045]\ \mathrm{GeV}$,
$M_{\mathrm{bc}} \in [5.27,\ 5.29] \ \mathrm{GeV}/c^2 $, and
$M_{1,2}(K\pi) \in [0.826,\ 0.966] \ \mathrm{GeV}/c^2$.

In about 17\% of events there are multiple 
$B^0 \to K^{*0}\kstarbar$ or $B^0 \to K^{*0}K^{*0}$ 
candidates. For these events we select the candidate with 
the smallest $\chi^2$ value for the $B^0$ decay vertex 
reconstruction. This selects the correct combination 87\% (97\%) 
of the time for longitudinally (transversely) polarized 
$B^0\to K^{*0} \kstarbar$ and $B^0\to K^{*0} K^{*0}$
decays.
The overall reconstruction efficiency for  
$B^0 \to K^{*0}\kstarbar$ as obtained from MC simulation is
4.43\% (5.23\%) for longitudinal (transverse) polarization. 
The overall reconstruction efficiency for
$B^0\to K^{*0} K^{*0}$ is 5.74\% (5.92\%) 
for longitudinal (transverse) polarization.
The efficiency for longitudinal polarization is lower,
as in this case the $K^{*0}$ daughters produce lower 
momentum kaons and pions.

The signal yields for $B^0 \to K^{*0}\kstarbar$ 
and other $B^0 \to K^+\pi^-K^-\pi^+$ decays
are extracted by performing an extended 
unbinned maximum likelihood (ML) fit to the variables
$M_{\mathrm{bc}}$, $\Delta E$, $M_1$, and $M_2$,
where $M_1\equiv M(K^+\pi^-)$ and 
$M_2\equiv M(K^-\pi^+)$. 
This four-dimensional fit discriminates among 
$K^{*0}\overline{K}{}^{*0}$, 
$K^{*0} K \pi$, 
$K_0^{*}(1430) \overline{K}{}_0^{*}(1430)$, 
$K_0^{*}(1430) \kstarbar$, 
$K_0^{*}(1430) K \pi$, and 
nonresonant $KK \pi\pi$ final states. 
Since there are large overlaps between these
states, we distinguish them by fitting 
a large $(M_1, M_2)$ region: 
$M_{1,2} \in [0.7,\ 1.7] \ \mathrm{GeV}/c^2$. 
We use a likelihood function 
\begin{equation}
\mathcal{L}= 
\exp \biggl(-\sum_{j} n_j \biggr)
\prod^{\mathrm{N_{cand}}}_{i=1}\biggl(\sum_{j} n_j P^i_j \biggr),
\end{equation}
where $i$ is the event identifier, 
$j$ indicates one of the event type 
categories for signals and backgrounds, 
$n_j$ denotes the yield of category~$j$, 
and $P^i_j$ is the probability density function (PDF)
of event~$i$ for category~$j$. The PDF is a product 
of two smoothed two-dimensional functions: 
$P^i_j =
P_j(M^i_{\mathrm{bc}}, \Delta E^i) \times P_j(M^i_1, M^i_2)
\equiv P_j(M^i_{\mathrm{bc}}, \Delta E^i, M^i_1, M^i_2)$.
The signal yields for $B^0 \to K^{*0} K^{*0}$ 
and other $B^0 \to K^+\pi^-K^+\pi^-$ decays are 
extracted by another four-dimensional fit in the 
same way, except that for this fit $M_2\equiv M(K^+\pi^-)$.

For the $B$ signal 
components, the smoothed functions
$P(M^{}_{\mathrm{bc}}, \Delta E^{})$ and $P(M^{}_1, M^{}_2)$ 
are obtained from MC simulation.
For the $M_{\mathrm{bc}}$ and $\Delta E$ PDFs, possible 
differences between data and the MC modeling
are calibrated using a large control sample of 
$B^0 \to D^-(K^+\pi^-\pi^-)\pi^+$ decays.
The signal mode PDF is divided into two parts: 
one is correctly reconstructed events (CR), and 
the other is ``self-cross-feed'' events (SCF) in which 
at least one track from the signal decay is replaced 
by one from the accompanying $B$ decay.
We use different PDFs for CR and SCF events
and fix the SCF fraction ($f_{\mathrm{SCF}}$)
to that obtained from MC simulation, i.e.,
\begin{eqnarray}
P^i_{\mathrm{Signal}} & = &
(1 - f^{}_{\mathrm{SCF}}) \times
P^{}_{\mathrm{CR}}(M^i_{\mathrm{bc}}, \Delta E^i, M^i_1, M^i_2) 
\nonumber  \\
 & & \ +\ f^{}_{\mathrm{SCF}} \times
P^{}_{\mathrm{SCF}}(M^i_{\mathrm{bc}}, \Delta E^i, M^i_1, M^i_2)\,.
\end{eqnarray}

For the continuum and $b \to c$ decay backgrounds, 
we use the product of a linear function for $\Delta E$, 
an ARGUS function~\cite{106} for $M_{\mathrm{bc}}$,
and a two-dimensional smoothed function for $M_1$-$M_2$.
The shape parameters of the linear and ARGUS functions 
for the continuum ($b \to c$) events are floated (fixed) 
in the fit; 
the shape of the $M_1$-$M_2$ functions for the continuum 
and $b \to c$ events are obtained from 
MC simulation 
and fixed in the fit.
The yields of the continuum and $b \to c$ decay backgrounds 
are floated in the fit.
For the charmless $B$ decay backgrounds, 
we use separate PDFs for 
$B^0\to K^+K^-K^0$, nonresonant $B^0\to K\pi\pi\pi$, 
and other charmless $B$ modes; 
all the PDFs 
are obtained from MC simulation. 
Note that the nonresonant $B^0\to K\pi\pi\pi$ decay will 
enter the sample if one of the pions is misidentified as 
a kaon; 
in this case the mean of the $\Delta E$ distribution 
shifts by about $+75$~MeV, 
since assigning a kaon mass instead of a pion mass 
increases the $B$ candidate energy. 
In the fit, we fix the yield of
$B^0\to K^+K^-K^0$ to 32 events, 
corresponding to a branching fraction of 
$24.7 \times10^{-6}$ \cite{PDGroup}, 
and the yield of other known charmless $B$ decays 
to that expected based on world average branching 
fractions~\cite{108}.
We set the branching fraction for $B^0\to K_2^{*}(1430)X$ 
to zero and only consider it in the systematics, 
as this mode has a large correlation
with $B^0\to K_{0}^{*}(1430) X$.
The yield of nonresonant $B^0\to K\pi\pi\pi$ is floated.
For the fully nonresonant modes, we assume the final-state 
particles are distributed uniformly 
in three- and four-body phase space. 

\begin{figure*}[htbp]
\centering
\epsfig{file=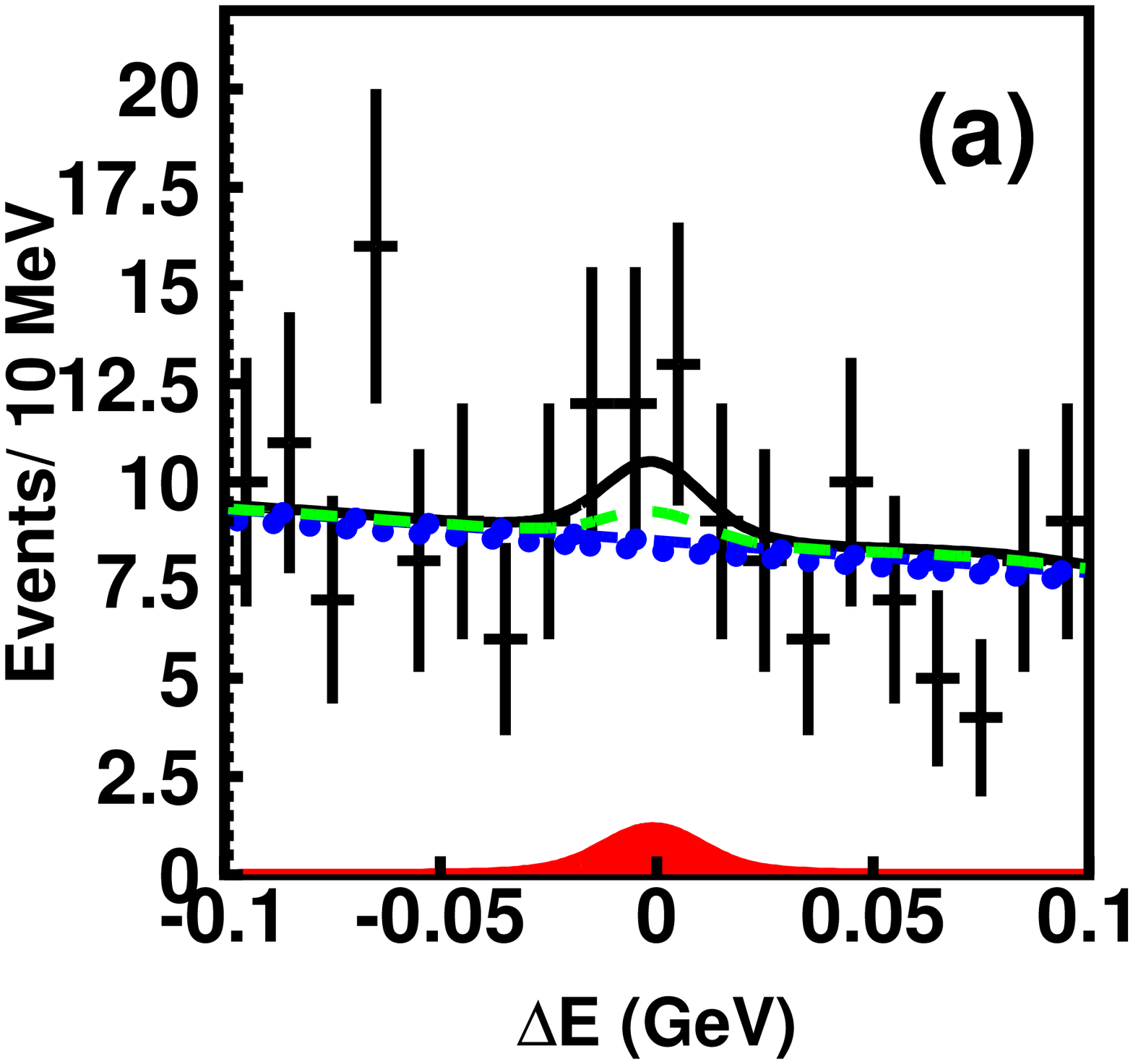,width=1.7in,height=1.7in}
\epsfig{file=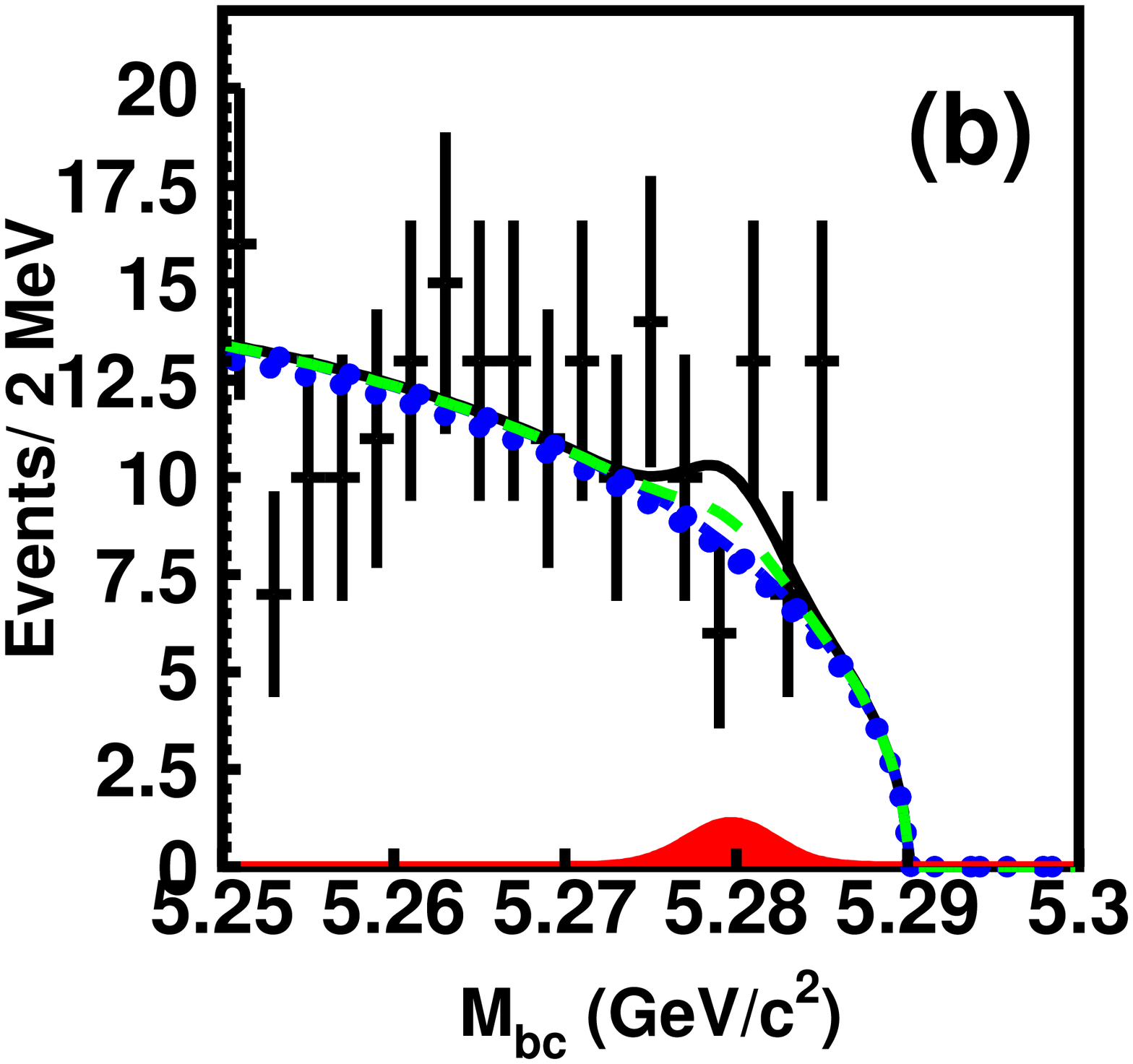,width=1.7in,height=1.7in}
\epsfig{file=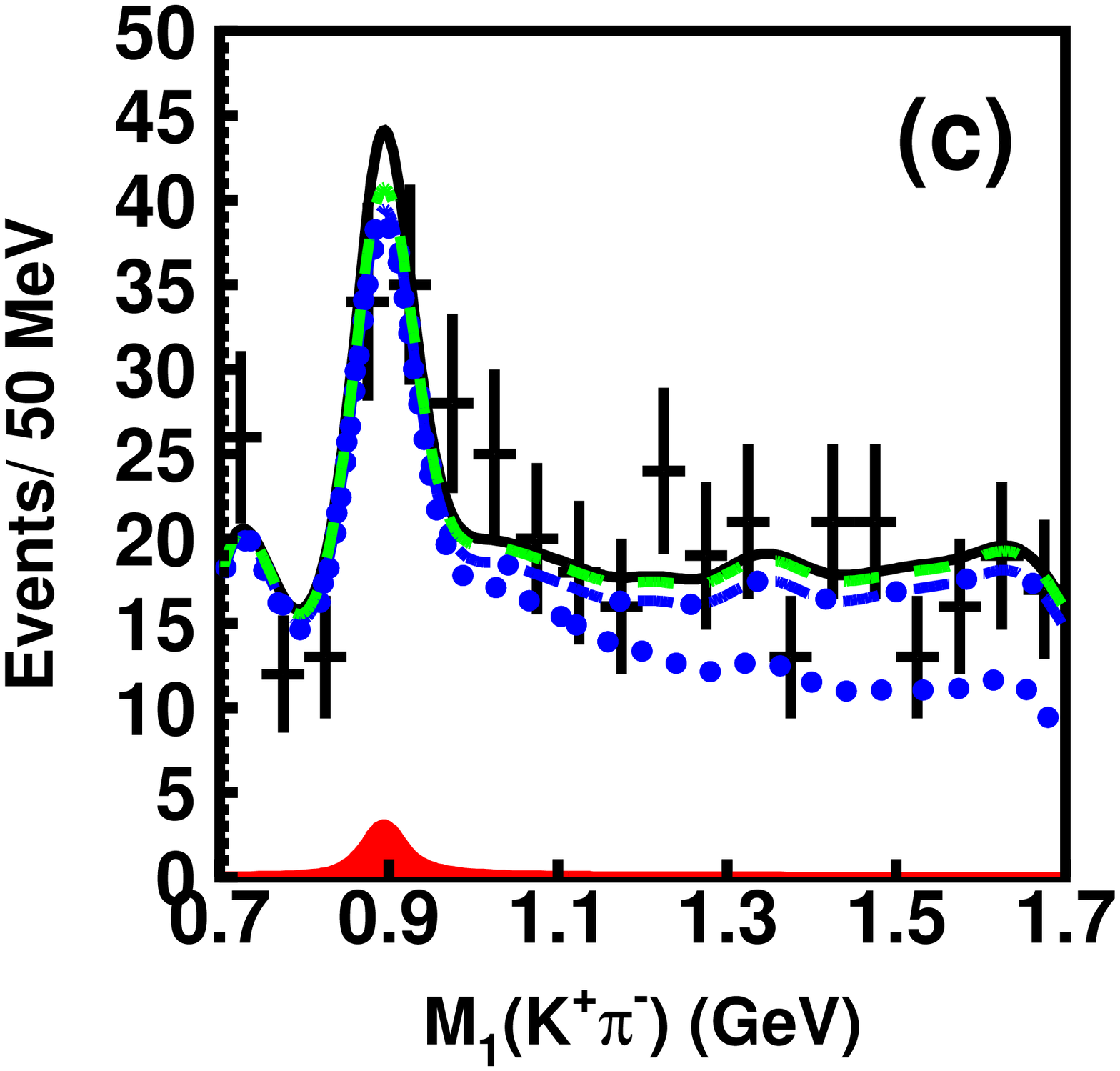,width=1.7in,height=1.7in}
\epsfig{file=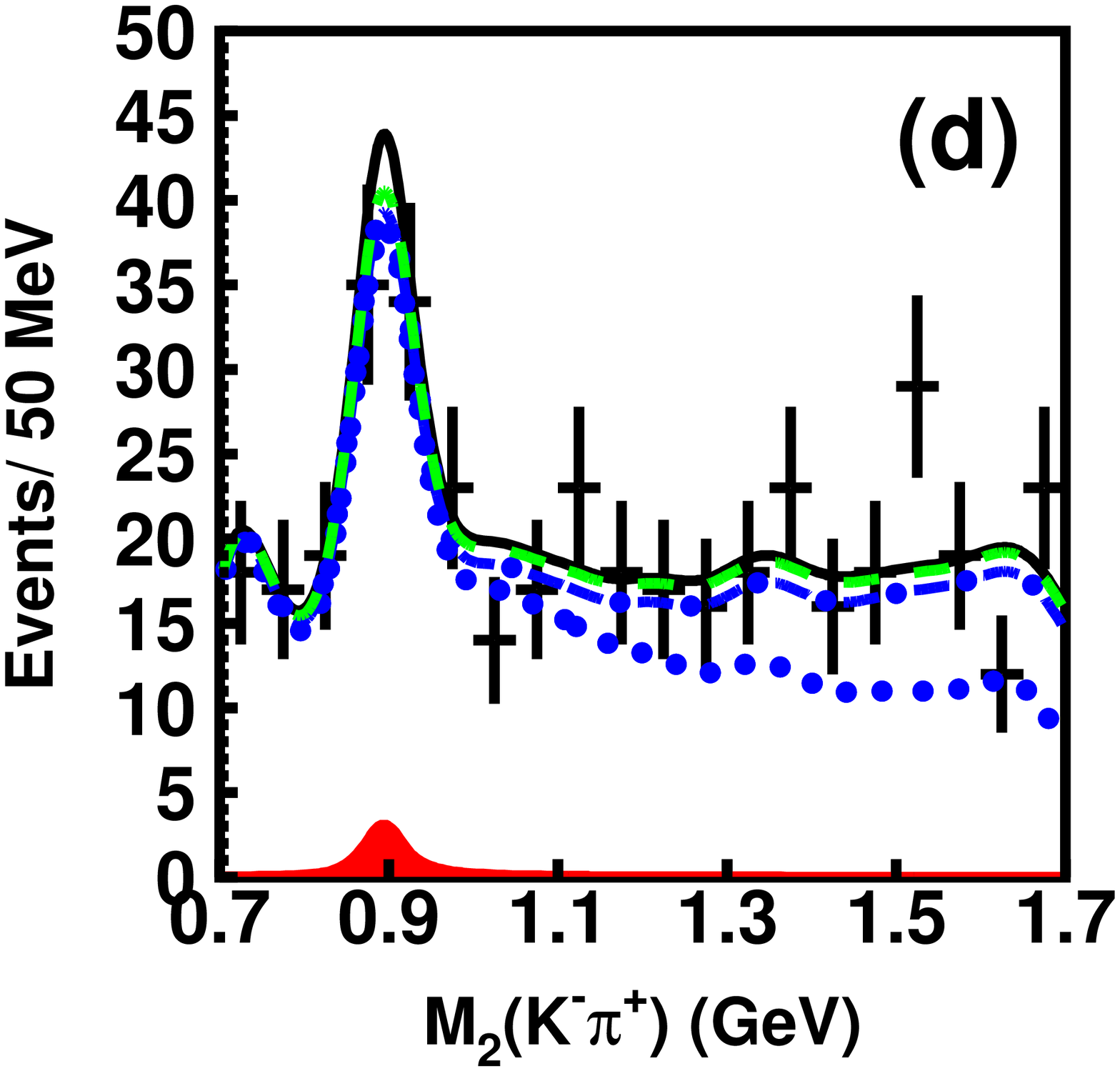,width=1.7in,height=1.7in}
\caption{
Projections of the four-dimensional fit onto (a) $\Delta E$, 
(b) $M_{\mathrm{bc}}$, (c) $M(K^+\pi^-)$, and 
(d) $M(K^-\pi^+)$
for candidates satisfying (except for the variable plotted) 
$\Delta E \in [-0.045,\ 0.045]\ \mathrm{GeV}$, 
$M_{\mathrm{bc}} \in [5.27,\ 5.29] \ \mathrm{GeV}/c^2 $, and 
$M_{1,2}(K\pi) \in [0.826,\ 0.966] \ \mathrm{GeV}/c^2$.
The thick solid curve shows the overall fit result; 
the solid shaded region represents the $B^0\to K^{*0} \kstarbar$ 
signal component; and the dotted, dot-dashed and dashed curves 
represent continuum background, $b\to c$ background, and 
charmless $B$ decay background, respectively.} 
\label{fig-fit1}
\end{figure*}

\begin{figure*}[htbp]
\centering
\epsfig{file=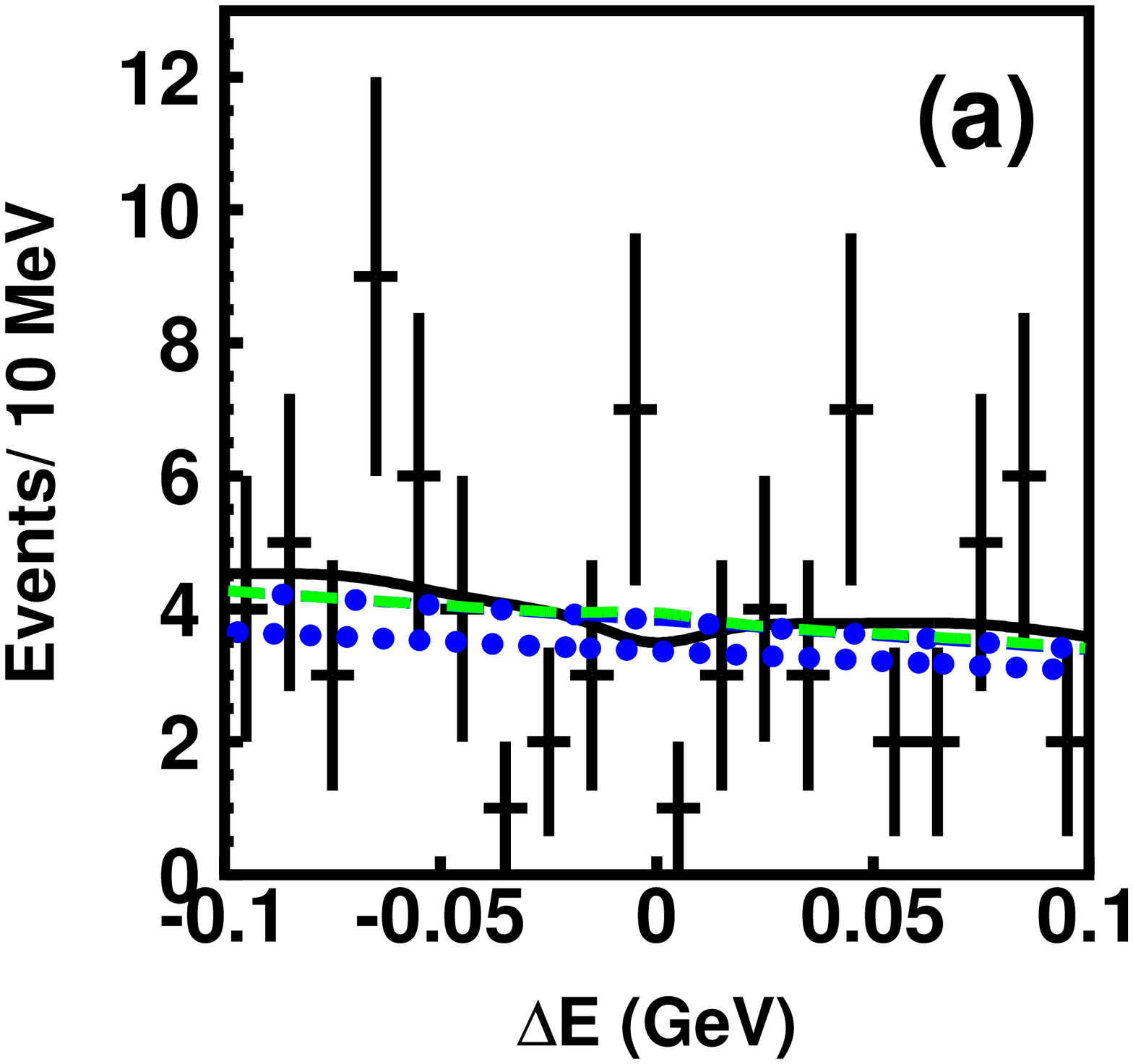,width=1.7in,height=1.7in}
\epsfig{file=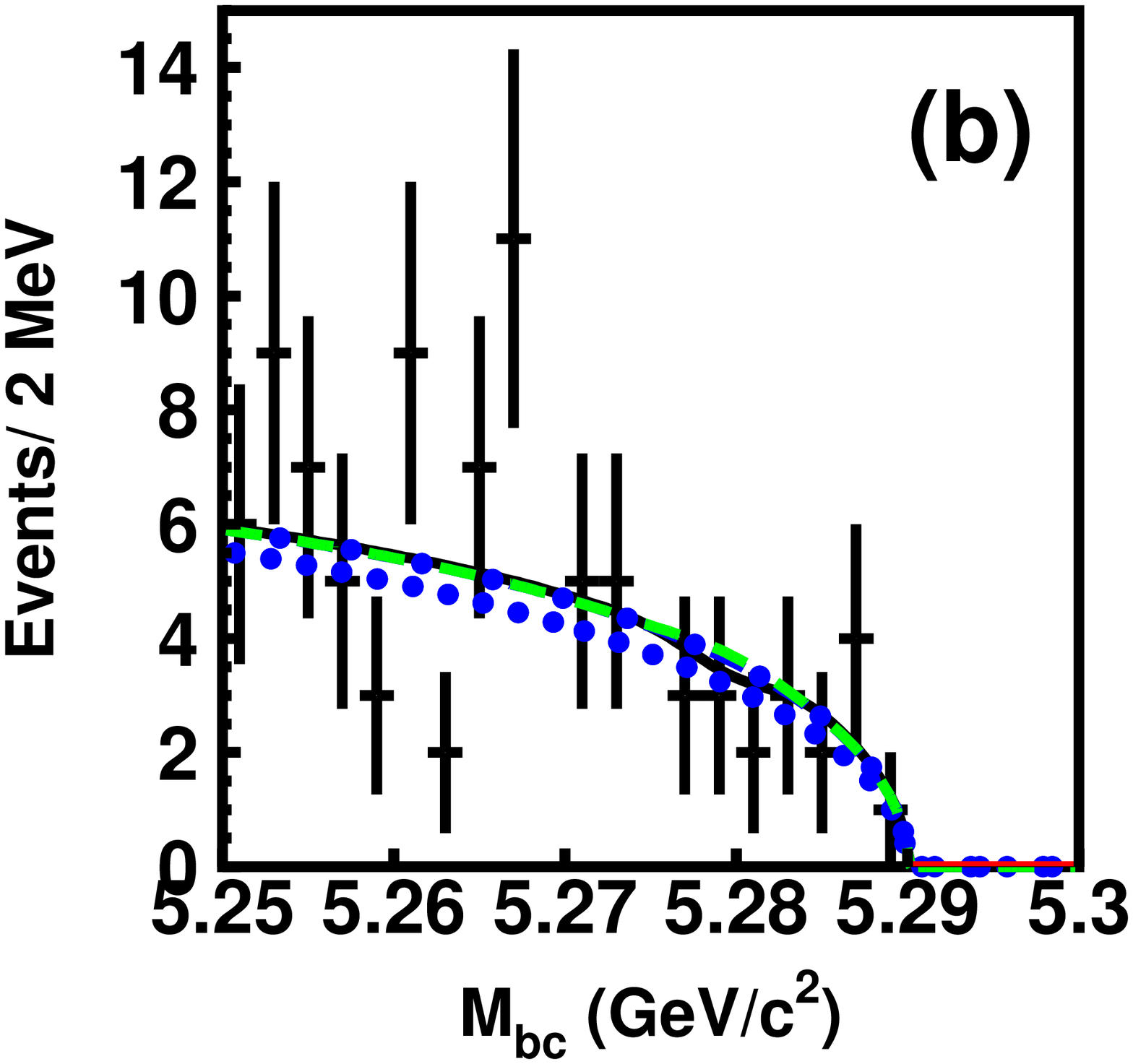,width=1.7in,height=1.7in}
\epsfig{file=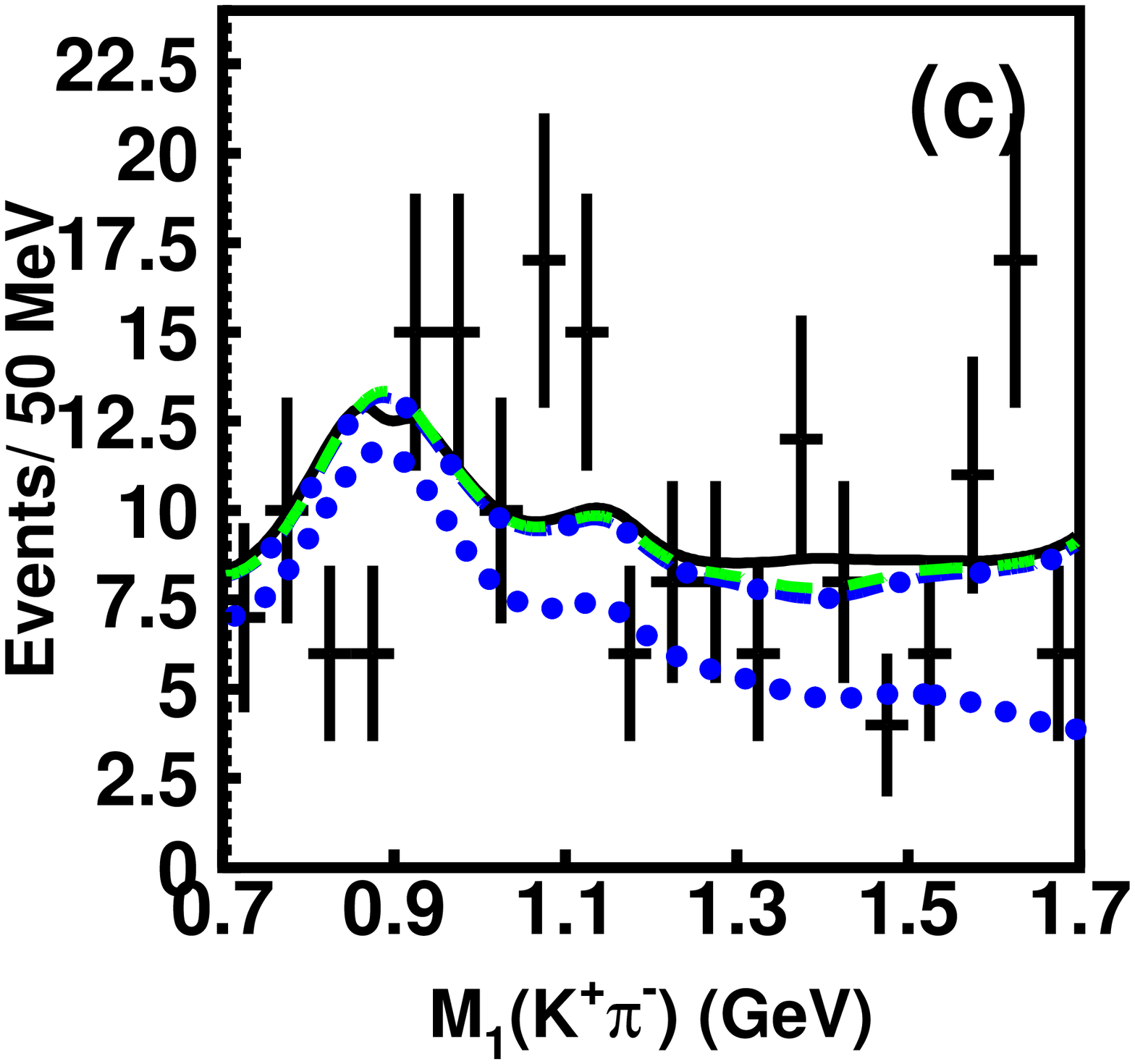,width=1.7in,height=1.7in}
\epsfig{file=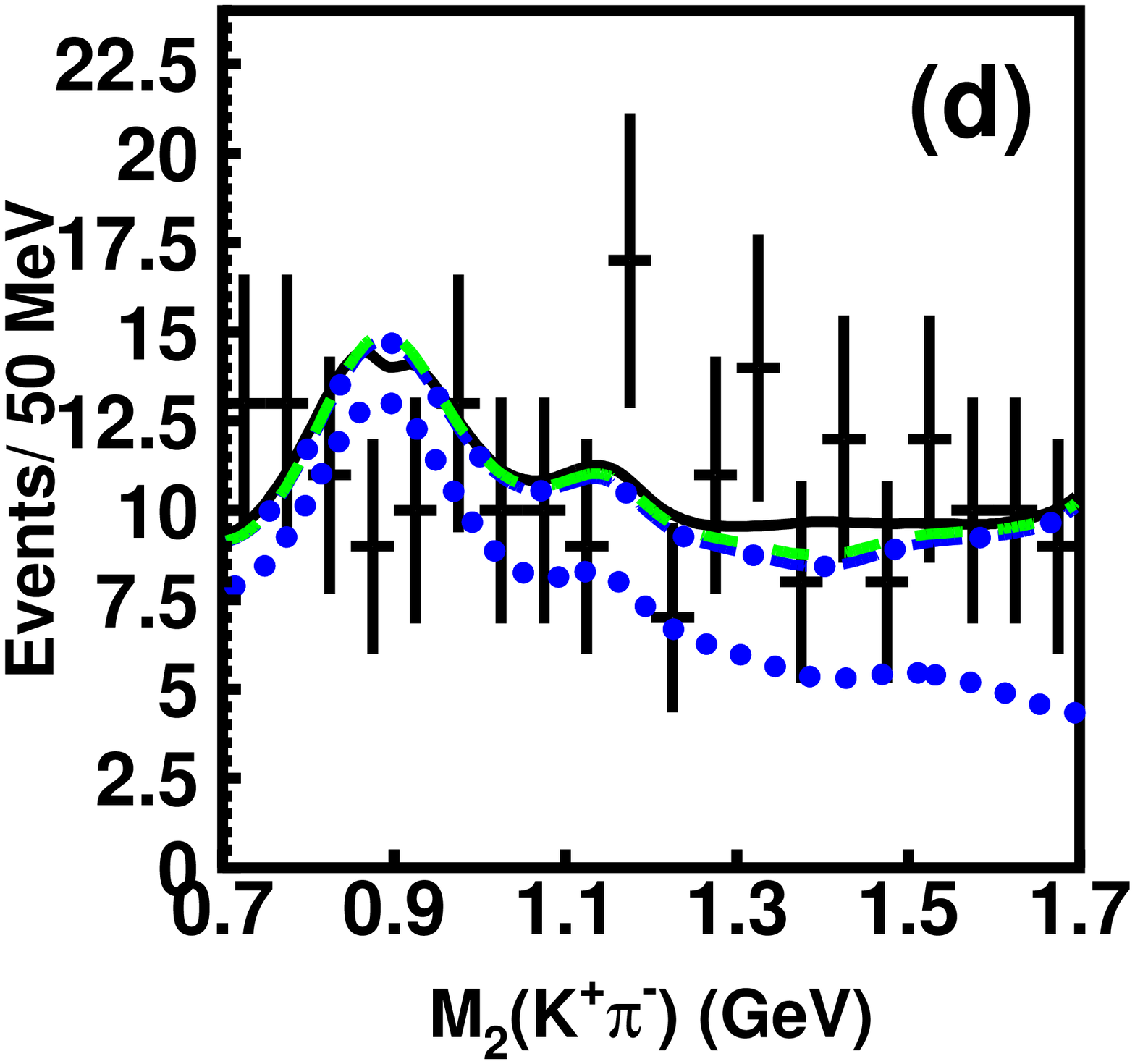,width=1.7in,height=1.7in}
\caption{
Same as for Fig.~\ref{fig-fit1} but for the 
$B^0\to K^{*0} K^{*0} \to (K^+\pi^-)(K^+\pi^-)$ study:
(a) $\Delta E$, (b) $M_{\mathrm{bc}}$, (c) $M_1(K^+\pi^-)$, and
(d) $M_2(K^+\pi^-)$.}
\label{fig-fit1-2}
\end{figure*}

The fit results are listed in Table~\ref{table-yield}, 
and projections of the fit superimposed 
to the data are shown in Figs.~\ref{fig-fit1} and \ref{fig-fit1-2}.
The statistical significance is calculated
as $\sqrt{-2\ln(\mathcal{L}_0 / \mathcal{L}_{\mathrm{max}})}$, 
where $\mathcal{L}_0$ and $\mathcal{L}_{\mathrm{max}}$ are the  
values of the likelihood function when the signal yield is 
fixed to zero and when it is allowed to vary, respectively. 
We do not find significant signals for $B^0\to K^{*0} \kstarbar$, 
$B^0\to K^{*0} K^{*0}$, and other charmless decay modes with 
$K^+\pi^- K^{\mp}\pi^{\pm}$ final states, 
and determine 90\% C.L. upper 
limits for the yields $(N)$. 
These limits are calculated via 
\begin{eqnarray}
  {{\int_0^N \mathcal{L}(x) dx} \over
  {\int_0^{\infty}\mathcal{L}(x) dx} } & = & 0.90\,,
\end{eqnarray}
where 
$x$ corresponds to the number of signal events. 
We include the systematic uncertainty in the upper limit (UL) 
by smearing the statistical likelihood function by a bifurcated 
Gaussian whose width is equal to the total systematic error. 
We also smear ${\cal L}$ when calculating the signal significance,
except that only the additive systematic errors related to 
signal yield are included in the convolved Gaussian width.
Our upper limits correspond to a longitudinal polarization 
fraction $f_{\mathrm{L}}=1$; as the efficiency for 
$f_{\mathrm{L}}<1$ is higher than that for 
$f_{\mathrm{L}}=1$, our limits are conservative.

\begin{table*}[htbp]
\renewcommand{\arraystretch}{1.4}
\begin{center}
\caption{Fit results for decay modes with final states 
$K^+\pi^-K^-\pi^+$ and $K^+\pi^-K^+\pi^-$.
The fit bias (in units of events) is obtained from MC simulation;
the yield includes the bias correction;
the efficiency $\varepsilon$ includes the 
PID efficiency correction and branching fractions for 
$K^{*0} \to K^+\pi^-$ and $K_0^{*}(1430) \to K^+\pi^-$
(66.5\% and 66.7\%, respectively); and the 
significance $\mathcal{S}$ is in units of $\sigma$. 
The first (second) error listed is statistical (systematic).
}
\begin{tabular}{lccccccccc}
\hline
\hline
Mode  
& Fit bias 
& Yield
& $\varepsilon$ (\%) 
& $\mathcal{S}$  
& $\mathcal{B}\times 10^6$ 
& UL $\times 10^6$ \cr \hline

$B^0\to K^{*0} \overline{K}{}^{*0}$
& $1.5 \pm 0.7$
& $7.7^{+9.7+2.8}_{-8.5-2.2}$
& 4.43 $(f_{\mathrm{L}} = 1.0)$
& 0.9
& $0.26^{+0.33+0.10}_{-0.29-0.08}$
& $< 0.8$ \cr

$B^0\to K^{*0} K^- \pi^+$
& $-5.4 \pm 2.9$
& $18.2^{+48.4+41.7}_{-45.3-40.9}$
& 1.31
& 0.3
& $2.11^{+5.63+4.85}_{-5.26-4.75}$
& $< 13.9$ \cr

$B^0\to K_0^{*}(1430) \overline{K}{}_0^{*}(1430)$
& $2.1 \pm 5.1$
& $78.5^{+70.6+56.4}_{-69.6-56.8}$
& 3.72
& 0.8
& $3.21^{+2.89+2.31}_{-2.85-2.32}$
& $< 8.4$ \cr

$B^0\to K_0^{*}(1430) \overline{K}{}^{*0}$
& $13.3 \pm 2.3$
& $19.6^{+31.1+40.0}_{-31.0-43.0}$
& 4.38
& 0.4
& $0.68 \pm 1.08 ^{+1.39}_{-1.49}$
& $< 3.3$ \cr

$B^0\to K_0^{*}(1430) K^- \pi^+$
& $14.6 \pm 9.8$
& $-222.8^{+171.5+159.8}_{-170.8-168.6}$
& 1.34
& ---
& ---
& $< 31.8$ \cr

Nonresonant $B^0\to K^+\pi^-K^-\pi^+$
& $-10.8 \pm 7.3$
& $158.4^{+120.6+104.1}_{-117.8-105.0}$
& 0.82
& 1.0
& $29.41^{+22.39+19.32}_{-21.87-19.49}$
& $< 71.7$ \cr 

\hline

$B^0\to K^{*0} K^{*0}$
& $1.0 \pm 0.5$
& $-3.7 \pm 3.3 ^{+2.5}_{-2.7}$
& 5.74 $(f_{\mathrm{L}} = 1.0)$
& ---
& ---
& $< 0.2$         \cr

$B^0\to K^{*0} K^+ \pi^-$
& $-2.5 \pm 2.7$
& $0.5 \pm 32.3 ^{+43.5}_{-40.1}$
& 1.93
& 0.0
& $0.04 \pm 2.55 ^{+3.43}_{-3.16}$
& $< 7.6$         \cr

$B^0\to K_0^{*}(1430) K_0^{*}(1430)$
& $3.4 \pm 1.3$
& $-28.4 \pm 16.1 ^{+87.7}_{-21.1}$
& 4.28
& ---
& ---
& $< 4.7$         \cr

$B^0\to K_0^{*}(1430) K^{*0}$
& $8.2 \pm 1.6$
& $8.0 \pm 18.7 ^{+23.9}_{-30.3}$
& 5.14
& 0.3
& $0.24 \pm 0.55 ^{+0.71}_{-0.90}$
& $< 1.7$      \cr

Nonresonant $B^0\to K^+\pi^-K^+\pi^-$
& $7.7 \pm 2.2$
& $10.8 \pm 28.3 ^{+31.4}_{-101.5}$
& 1.98
& 0.3
& $0.83 \pm 2.17 ^{+2.42}_{-7.80}$
& $< 6.0$     \cr

\hline \hline
\end{tabular}
\label{table-yield}
\end{center}
\end{table*}

To check our reconstruction efficiencies, we measure the 
yields of control samples
$B^0\to D^{-}\pi^{+} \to (K^+K^-\pi^{-})\pi^{+}$ 
and 
$B^0\to \dbar K^{*0} \to (K^+\pi^-)(K^+\pi^-)$.
These modes have a similar topology to the signal modes
and are selected using the same selection criteria
except that, instead of $D$ vetos, we require
$|M(KK\pi)-m_{D^{\pm}}|<13 \ \mathrm{MeV}/c^2$, 
$|M(K\pi)-m_{D^0}|<13 \ \mathrm{MeV}/c^2$, and 
$826~\mathrm{MeV}/c^2<M(K\pi)<966~\mathrm{MeV}/c^2$. 
The efficiencies are 19\% for $B^0\to D^{-}\pi^{+}$
and 11\% for $B^0\to \dbar K^{*0}$. The difference
in signal yields between the measured and 
expected values
are $(5.8 \pm 5.8)$\% and $(5.6 \pm 27.8)$\% for
$B^0\to D^{-}\pi^{+}$ and $B^0\to \dbar K^{*0}$, respectively.
These differences are consistent with zero.
 
The systematic errors (in units of events) 
are summarized in Tables~\ref{table-sys-sum} 
and~\ref{table-sys-sum-2}.
For systematic uncertainties due to fixed yields,
e.g., that of charmless $B$ background, 
we vary the yields by their 
uncertainties ($\pm 1\sigma$).
For the systematic uncertainties due to
$B^0\to K_2^{*}(1430)X$ decays, including
$B^0\to K_2^{*}(1430) \overline{K}{}_2^{*}(1430)$,
$B^0\to K_2^{*}(1430) \overline{K}{}_0^{*}(1430)$,
$B^0\to K_2^{*}(1430) \kstarbar$, and
$B^0\to K_2^{*}(1430) K \pi$,
we float their yields in the four-dimensional ML fit;
the differences between these results and the nominal 
fit values are taken as systematic errors. Systematic 
uncertainties for the $\Delta E$-$M_{\rm bc}$ PDFs 
are estimated by varying the signal peak positions 
and resolutions by $\pm 1\sigma$ and repeating the fits.
Systematic uncertainties for the $M_1$-$M_2$ PDFs
are estimated in a similar way;
we vary the mean and width of 
$K^{*0}$ and $K_0^{*}(1430)$ 
mass shapes according to 
the uncertainties in the world average values~\cite{PDGroup}.
A systematic error for the longitudinal polarization fraction 
is obtained by changing the fraction from the nominal 
value $f_{\mathrm{L}}\!=\!1$ to the lowest possible value
$f_{\mathrm{L}}\!=\!0$ when evaluating the reconstruction efficiency.
According to MC simulation, the signal SCF fractions are
13.4\% for (longitudinally polarized) $B^0\to K^{*0} \kstarbar$,
7.9\% for $B^0\to K^{*0} K \pi$,
6.7\% for $B^0\to K_0^{*}(1430) \overline{K}{}_0^{*}(1430)$,
6.7\% for $B^0\to K_0^{*}(1430) \kstarbar$,
7.6\% for $B^0\to K_0^{*}(1430) K \pi$, and 
9.2\% for nonresonant $B^0\to KK\pi\pi$.
We estimate a systematic uncertainty due to
these fractions
by varying them by $\pm 50\%$.

A high-statistics MC study indicates that there 
are small fit biases; 
these are listed in 
Table~\ref{table-yield}.
We find that fit biases occur due to the correlations between the
two sets of variables ($\Delta E$, $M_{\rm bc}$) and ($M_1$, $M_2$),
which are not taken into account in our fit.
We correct the fitted yields 
for these biases.
To take into account possible differences 
between 
MC simulation 
and data, we take both the
magnitude of the bias corrections and the
uncertainty in the corrections as systematic
errors. The systematic errors for the efficiency 
arise from the tracking efficiency, PID, and
the $\mathcal{R}$ requirement.
The systematic error on the track-finding efficiency
is estimated to be 1.2\% per track using partially
reconstructed $D^*$ events. The systematic error due 
to the PID is 1.0\% per track as estimated using an 
inclusive $D^*$ control sample.
The systematic error for the $\mathcal{R}$ requirement
is determined from the efficiency difference 
between data and MC samples of
$B^0 \to D^-(K^+\pi^-\pi^-)\pi^+$ decays.

\begin{table*}[htbp]
\begin{center}
\caption{
Summary of systematic errors (in units of events)
for decay modes with a final state $K^+\pi^-K^-\pi^+$.
The parameters 
$N_{B^0\to K_2^{*}(1430)X}$ and $N_{b\to u,d,s}$
(the other known charmless $B$ decays)
correspond to branching fraction uncertainties for these charmless 
$B$ decays. Values for $f_{\mathrm{L}}$ and $f_{\mathrm{SCF}}$ 
are the uncertainties for longitudinal polarization 
and self-cross-feed, respectively.}
\begin{tabular}{lccccccccc}
\hline
\hline

Source & $K^{*0} \kstarbar$
       & $K^{*0} K^- \pi^+$
       & $K_0^{*}(1430) \overline{K}{}_0^{*}(1430)$
       & $K_0^{*}(1430) \overline{K}{}^{*0}$
       & $K_0^{*}(1430) K^- \pi^+$
       & Nonresonant $K^+\pi^-K^-\pi^+$       \cr \hline

Fitting PDF 
& $\pm 1.8$ 
& $\pm 40.3$ 
& $\pm 55.6$ 
& $\pm 37.7$ 
& $\pm 158.0$ 
& $\pm 102.4$ \cr

$N_{B^0\to K_2^{*}(1430)X}$ 
& $+1.1$ 
& $+10.2$ 
& $-7.1$ 
& $-20.4$ 
& $-52.8$ 
& $-10.5$  \cr

$N_{b\to u,d,s}$
& $\pm 0.0$ 
& $\pm 0.1$ 
& $\pm 0.2$ 
& $\pm 0.1$ 
& $\pm 0.7$ 
& $\pm 1.0$  \cr

$f_{\mathrm{L}}$ 
& $-0.1$ & --- & --- & --- & --- & --- \cr

$f_{\mathrm{SCF}}$ 
& $\pm 0.7$ 
& $\pm 1.4$ 
& $\pm 5.6$ 
& $\pm 2.2$ 
& $\pm 17.2$ 
& $\pm 14.6$ \cr

Fit Bias 
& $^{+1.5}_{-0.7}$ 
& $^{+2.9}_{-5.4}$ 
& $\pm 5.1$ 
& $^{+13.3}_{-2.3}$ 
& $^{+8.9}_{-14.7}$ 
& $^{+7.3}_{-10.8}$ \cr

Tracking 
& $\pm 0.4$ 
& $\pm 0.8$ 
& $\pm 3.5$ 
& $\pm 0.9$ 
& $\pm 9.6$ 
& $\pm 6.8$ \cr

PID 
& $\pm 0.4$ 
& $\pm 0.7$ 
& $\pm 2.9$ 
& $\pm 0.7$ 
& $\pm 8.2$ 
& $\pm 6.0$  \cr

$\mathcal{R}$ requirement 
& $\pm 0.2$ 
& $\pm 0.4$ 
& $\pm 1.6$ 
& $\pm 0.4$ 
& $\pm 4.5$ 
& $\pm 3.2$  \cr

$N_{B \overline B}$ 
& $\pm 0.1$ 
& $\pm 0.3$ 
& $\pm 1.1$ 
& $\pm 0.3$ 
& $\pm 3.1$ 
& $\pm 2.2$  \cr\hline

Sum 
& $^{+2.8}_{-2.1}$ 
& $^{+41.7}_{-40.7}$ 
& $^{+56.3}_{-56.8}$ 
& $^{+40.0}_{-43.0}$ 
& $^{+159.7}_{-168.6}$
& $^{+104.1}_{-104.9}$ \cr

\hline
\hline
\end{tabular}
\label{table-sys-sum}
\end{center}
\end{table*}

\begin{table*}[htbp]
\begin{center}
\caption{
Same as for Table~\ref{table-sys-sum}
but for decay modes with a final state
$K^+\pi^-K^+\pi^-$. 
}
\begin{tabular}{lccccccccc}
\hline
\hline

Source & $K^{*0} K^{*0}$
       & $K^{*0} K^+ \pi^-$
       & $K_0^{*}(1430) K_0^{*}(1430)$
       & $K_0^{*}(1430) K^{*0}$
       & Nonresonant $K^+\pi^-K^+\pi^-$
       \cr \hline

Fitting PDF 
& $\pm 2.4$ 
& $\pm 40.0$ 
& $\pm 20.7$ 
& $\pm 22.5$ 
& $\pm 30.5$ \cr

$N_{B^0\to K_2^{*}(1430)X,K_0^{*}(1430)K\pi}$ 
& $-0.3$ 
& $+16.8$ 
& $+85.1$ 
& $-20.2$ 
& $-96.8$ \cr

$N_{b\to u,d,s}$
& $\pm 0.0$ 
& $\pm 0.4$ 
& $\pm 0.1$ 
& $\pm 0.3$ 
& $\pm 0.3$ \cr

$f_{\mathrm{L}}$ 
& $-0.7$ & --- & --- & --- & --- \cr

$f_{\mathrm{SCF}}$ 
& $\pm 0.1$ 
& $\pm 0.2$ 
& $\pm 0.8$ 
& $\pm 0.5$ 
& $\pm 0.2$ \cr

Fit Bias 
& $^{+0.5}_{-1.0}$ 
& $\pm 2.7$ 
& $^{+1.3}_{-3.4}$ 
& $^{+8.2}_{-1.6}$ 
& $^{+7.7}_{-2.2}$ \cr

Tracking 
& $\pm 0.2$ 
& $\pm 0.0$ 
& $\pm 1.2$ 
& $\pm 0.4$ 
& $\pm 0.5$ \cr

PID 
& $\pm 0.2$ 
& $\pm 0.0$ 
& $\pm 1.0$ 
& $\pm 0.3$ 
& $\pm 0.4$ \cr

$\mathcal{R}$ requirement 
& $\pm 0.1$ 
& $\pm 0.0$ 
& $\pm 0.6$ 
& $\pm 0.2$ 
& $\pm 0.2$ \cr

$N_{B \overline B}$ 
& $\pm 0.1$ 
& $\pm 0.0$ 
& $\pm 0.4$ 
& $\pm 0.1$ 
& $\pm 0.2$ \cr\hline

Sum 
& $^{+2.5}_{-2.7}$ 
& $^{+43.5}_{-40.1}$ 
& $^{+87.7}_{-21.1}$ 
& $^{+23.9}_{-30.3}$
& $^{+31.4}_{-101.5}$ \cr

\hline
\hline
\end{tabular}
\label{table-sys-sum-2}
\end{center}
\end{table*}

In summary, we have used a 
data sample corresponding to 
$657 \times 10^6$ $B \overline B$ pairs
to search for $B^0\to K^{*0} \kstarbar$, 
$B^0\to K^{*0} K^{*0}$, and other charmless 
decay modes with a $K^+\pi^- K^{\mp}\pi^{\pm}$ final state.
We do not find significant signals for any of 
these modes. Our measured branching fraction for 
$B^0\to K^{*0} \kstarbar$ is 
$(0.26^{+0.33+0.10}_{-0.29-0.07})\times 10^{-6}$, 
which is lower than that obtained by 
BaBar \cite{aaubert} by 2.2$\sigma$. 
Our 90\% C.L. upper limits are $0.8 \times 10^{-6}$ 
for $\mathcal{B}(B^0\to K^{*0} \kstarbar)$ and 
$0.2 \times 10^{-6}$ for $\mathcal{B}(B^0\to K^{*0} K^{*0})$; 
those for other decay modes are listed in 
Table~\ref{table-yield}. 

We thank the KEKB group for excellent operation of the
accelerator, the KEK cryogenics group for efficient solenoid
operations, and the KEK computer group and
the NII for valuable computing and SINET3 network support.  
We acknowledge support from MEXT, JSPS and Nagoya's TLPRC (Japan);
ARC and DIISR (Australia); NSFC (China); MSMT (Czechia);
DST (India); MEST, NRF, NSDC of KISTI (Korea); MNiSW (Poland); 
MES and RFAAE (Russia); ARRS (Slovenia); SNSF (Switzerland); 
NSC and MOE (Taiwan); and DOE (USA).


\begin{thebibliography}{99}
\bibitem{chargeconjugates}
Charge-conjugate modes are implicitly included 
throughout this paper unless noted otherwise.
\end{thebibliography}
\end{document}